\documentclass[english]{article}
\usepackage[T1]{fontenc}
\usepackage[latin9]{inputenc}
\usepackage[letterpaper]{geometry}
\usepackage{amsmath}
\usepackage{graphicx}
\usepackage{amssymb}
\usepackage{esint}

\makeatletter

\newcommand{\lyxmathsym}[1]{\ifmmode\begingroup\def\b@ld{bold}
  \text{\ifx\math@version\b@ld\bfseries\fi#1}\endgroup\else#1\fi}

\providecommand{\tabularnewline}{\\}

\makeatother

\usepackage{babel}

\begin{document}

\title{Milli-interacting Dark Matter}

\author{Quentin Wallemacq%
\thanks{quentin.wallemacq@ulg.ac.be%
}}

\maketitle
\begin{center}
IFPA, D\'ep. AGO, Universit\'e de Li\`ege, Sart Tilman, 4000 Li\`ege, Belgium
\par\end{center}
\begin{abstract}
We present a dark matter model reproducing well the results from DAMA/LIBRA
and CoGeNT and having no contradiction with the negative results from
XENON100 and CDMS-II/Ge. Two new species of fermions $F$ and $G$
form hydrogen-like atoms with standard atomic size through a dark
$U\left(1\right)$ gauge interaction carried out by a dark massless
photon. A Yukawa coupling between the nuclei $F$ and neutral scalar
particles $S$ induces an attractive shorter-range interaction. This
dark sector interacts with our standard particles because of the presence
of two mixings, a kinetic photon - dark photon mixing, and a mass
$\sigma-S$ mixing. The dark atoms from the halo diffuse elastically
in terrestrial matter until they thermalize and then reach underground
detectors with thermal energies, where they form bound states with
nuclei by radiative capture. This causes the emission of photons that
produce the signals observed by direct-search experiments.
\end{abstract}

\section{Introduction}

Direct searches for dark matter have been accumulating results in
recent years, starting with the DAMA/NaI experiment that observed
a significant signal since the late 90's. Its successor, DAMA/LIBRA,
has further confirmed the signal and improved its statistical significance
to a current value of $8.9\,\sigma$ \cite{Bernabei:2010mq}. Some
other experiments such as CoGeNT \cite{Aalseth:2012if}, CRESST-II
\cite{Angloher:2011uu}, and very recently CDMS-II/Si \cite{Agnese:2013rvf},
are going in the same direction and report observations of events
in their underground detectors, while others, such as XENON100 \cite{Aprile:2012nq},
or CDMS-II/Ge \cite{Ahmed:2010wy} continue to rule out any detection.
These experiments challenge the usual interpretation of dark matter
as being made only of Weakly Interacting Massive Particles (WIMPs).
Because of the motion of the solar system in the galactic dark-matter
halo, incident WIMPs would hit underground detectors where they could
produce nuclear recoils, which would then be the source of the observed
signals. However, this interpretation of the data induces strong contradictions
between experiments with positive and negative results as well as
tensions between experiments with positive results \cite{Aprile:2012nq}.

In this context, alternatives have been proposed to reconcile the
experiments. Among them, mirror matter \cite{Foot:2012rk} and millicharged
atomic dark matter \cite{Cline:2012is} provide explanations respectively
in terms of Coulomb scattering of millicharged mirror nuclei on nuclei
in the detectors or in terms of hyperfine transitions of millicharged
dark atoms analogous to hydrogen colliding on nuclei. In these scenarios,
millicharged dark species are obtained by a kinetic mixing between
standard photons and photons from the dark sector. Mirror matter in
the presence of kinetic photon - mirror photon mixing gives a rich
phenomenology that can reproduce the signals of most of the experiments,
but some tensions remain with experiments such as XENON100 or EDELWEISS.
Millicharged atomic dark matter can explain the excess of events reported
by CoGeNT but keeps the contradictions with the others.

Another scenario has been proposed by Khlopov $et\, al.$ \cite{Khlopov:2010ik,Khlopov:2011me},
in which new negatively charged particles (O$^{--}$) are bound to
primordial helium (He$^{++}$) in neutral O-helium dark atoms (OHe).
The approach here is quite different because the interactions of these
OHe with terrestrial matter are determined by the nuclear interactions
of the helium component. Therefore, instead of producing nuclear recoils,
these dark atoms would thermalize in the Earth by elastic collisions
and reach underground detectors with thermal energies, where they
form bound states with nuclei by radiative capture, the emitted photon
being the source of the signal. Therefore, the observation of a signal
depends on the existence of a bound state in the OHe - nucleus system
and can provide a natural explanation to the negative results experiments,
in case of the absence of bound states with the constituent nuclei.
However, a careful analysis of the interactions of OHe atoms with
nuclei \cite{Cudell:2012fw} has ruled out the model. Nevertheless,
the scenario presented here keeps many of the features of the OHe,
but avoids its problems.

Our model aims at solving the discrepancies between experiments with
positive results, as well as to reconcile them with those without
any signal. It presents common features with the ones mentioned above
\cite{Foot:2012rk,Cline:2012is,Khlopov:2010ik}. It contains dark
fermions that possess electric millicharges due to the same kind of
photon - dark photon mixing as in the mirror and atomic-dark-matter
scenarios, but also another mixing between $\sigma$ mesons and new
dark-scalar particles creating an attractive interaction with nucleons,
which couple to $\sigma$ mesons in the framework of an effective
Yukawa theory. The dark matter will be in the form of hydrogenoid
atoms with standard atomic sizes that interact sufficiently with terrestrial
matter to thermalize before reaching underground detectors. There,
dark and standard nuclei will form bound states by radiative capture
through the attractive exchange between dark fermions and nuclei.

An important feature of such a model is that it presents a self-interacting
dark matter, on which constraints exist from the Bullet cluster or
from halo shapes \cite{2002ApJ...564...60M}. According to \cite{Fan:2013yva},
these can be avoided if the self-interacting candidate is reduced
to at most $5\%$ of the dark matter mass content of the galaxy, the
rest being constituted by conventional collisionless particles. In
the following, the dark sector will therefore be a subdominant part
of dark matter.

In Section \ref{sec:The-model}, the ingredients and the effective
lagrangian of the model are described. Constraints from vector-meson
disintegrations are considered and the interaction potentials between
dark and standard sectors are derived in Section \ref{sec:Dark-standard-interactions},
from the lagrangian of Section \ref{sec:The-model}. The thermalization
of the dark atoms in terrestrial matter is studied in Section \ref{sec:Thermalization-of-dark}
and constraints on model parameters are obtained, to thermalize between
the surface and an underground detector. The radiative-capture process
within a detector is described in Section \ref{sec:Interactions-in-underground},
where the capture cross section and the event rate are derived. Section
\ref{sec:Results} gives an overview of the reproduction of the experimental
results.

\section{The model\label{sec:The-model}}

We postulate that a dark, hidden, sector exists, consisting of two
kinds of new fermions, denoted by $F$ and $G$, respectively coupled
to dark photons $\Gamma$ with opposite couplings $+e'$ and $-e'$,
while only $F$ is coupled to neutral dark scalars $S$ with a positive
coupling $g'$. This dark sector is governed by the lagrangian \begin{equation}
\mathcal{L}^{dark}=\mathcal{L}_{0}^{dark}+\mathcal{L}_{int}^{dark}\label{eq:1}\end{equation}
 where the free and interaction lagrangians $\mathcal{L}_{0}^{dark}$
and $\mathcal{L}_{int}^{dark}$ have the forms \begin{equation}
\mathcal{L}_{0}^{dark}=\sum_{k=F,G}\overline{\psi}_{k}\left(i\gamma^{\mu}\partial_{\mu}-m_{k}\right)\psi_{k}-\frac{1}{4}F'^{\mu\nu}F'_{\mu\nu}+\frac{1}{2}\partial_{\mu}\phi_{S}\partial^{\mu}\phi_{S}-\frac{1}{2}m_{S}\phi_{S}^{2}\label{eq:2}\end{equation}
 and \begin{equation}
\mathcal{L}_{int}^{dark}=e'\overline{\psi}_{F}\gamma^{\mu}A'_{\mu}\psi_{F}-e'\overline{\psi}_{G}\gamma^{\mu}A'_{\mu}\psi_{G}+g'\phi_{S}\overline{\psi}_{F}\psi_{F}\label{eq:3}\end{equation}
Here, $\psi_{F(G)}$ , $A'$ and $\phi_{S}$ are respectively the
fermionic, vectorial and real scalar fields of the dark fermion $F(G)$
, dark photon $\Gamma$ and dark scalar $S$, while $m_{F(G)}$ and
$m_{S}$ are the masses of the $F(G)$ and $S$ particles. $F'$ stands
for the electromagnetic-field-strength tensor of the massless dark
photon $\Gamma$.

Moreover, we assume that the dark photons $\Gamma$ and the dark scalars
$S$ are mixed respectively with the standard photons $\gamma$ and
neutral mesons $\sigma$ through the mixing lagrangian \begin{equation}
\mathcal{L}_{mix}=\frac{1}{2}\tilde{\epsilon}F^{\mu\nu}F'_{\mu\nu}+\tilde{\eta}\left(m_{\sigma}^{2}+m_{S}^{2}\right)\phi_{\sigma}\phi_{S}\label{eq:4}\end{equation}
 where $m_{\sigma}=600$ MeV \cite{Amsler:2008zzb} is the mass of
$\sigma$ and $\tilde{\epsilon}$ and $\tilde{\eta}$ are the dimensionless
parameters of kinetic $\gamma-\Gamma$ and mass $\sigma-S$ mixings.
These are supposed to be small compared with unity.

The model therefore contains $7$ free parameters, $m_{F}$, $m_{G}$,
$m_{S}$, $e'$, $g'$, $\tilde{\epsilon}$ and $\tilde{\eta}$, and
the total lagrangian of the combined standard and dark sectors is
\begin{equation}
\mathcal{L}=\mathcal{L}^{SM}+\mathcal{L}^{dark}+\mathcal{L}_{mix}\label{eq:26}\end{equation}
 where $\mathcal{L}^{SM}$ stands for the lagrangian of the standard
model.

The $F$ and $G$ fermions will form dark hydrogenoid atoms in which
$F$ will play the role of a dark nucleus binding to nuclei in underground
detectors, while $G$ acts as a dark electron. $F$ has then to be
heavy enough to form bound states and we will seek masses of $F$
between $10$ GeV and $10$ TeV, while requiring $m_{G}\ll m_{F}$.
Due to the mass mixing term in \eqref{eq:4}, $F$ will interact with
nucleons through the exchange of $S$ and this attractive interaction
will be responsible for the binding. It cannot be too long-ranged
but it must allow the existence of nucleus - $F$ bound states of
at least the size of the nucleus. Because the range of the interaction
is of the order of $m_{S}^{-1}$, this leads us to consider values
of the mass of $S$ between $100$ keV and $10$ MeV. The other $4$
parameters will not be directly constrained by the direct-search experiments,
but only the products $\tilde{\epsilon}e'$ and $\tilde{\eta}g'$.
However, a reasonable choice seems to be $\tilde{\epsilon},\tilde{\eta}\ll1$
together with $e'\simeq e$ and $g'\simeq g$, where $e$ is the charge
of the proton and $g=14.4$ \cite{Erkol:2005jz} is the Yukawa coupling
of the nucleon to the $\sigma$ meson. In summary, we will consider
:\[
    \left \{
    \begin{array}{c}
       10 $ GeV $ \leq\ m_F \leq\ 10 $ TeV$ \\
       100 $ keV $ \leq\ m_S \leq\ 10 $ MeV$ \\
       m_G \ll\ m_F \\
       e' \simeq\ e \\
       g' \simeq\ g \\
       \tilde \epsilon,\tilde \eta \ll\ 1
    \end{array}
    \right .
\]

\section{Dark-standard interactions\label{sec:Dark-standard-interactions}}

The mixings described by \eqref{eq:4} induce interactions \cite{Foot:2012rk,Cline:2012is}
between dark fermions $F$ and $G$ and our standard particles. It
is well known that, to first order in $\tilde{\epsilon}$, a kinetic
mixing such as the one present in \eqref{eq:4} will make the dark
particles $F$ and $G$ acquire small effective couplings $\pm\tilde{\epsilon}e'$
to the standard photons. One can define the kinetic mixing parameter
in terms of the electric charge of the proton $e$ through $\epsilon e\equiv\tilde{\epsilon}e'$,
which means that the particles $F$ and $G$ will interact electromagnetically
with any charged particle of the standard model with millicharges
$\pm\epsilon e$.

The mass mixing from \eqref{eq:4} characterized by $\tilde{\eta}$
induces an interaction between $F$ and $\sigma$, through the exchange
of $S$, and hence an interaction between $F$ and any standard particle
coupled to $\sigma$, e.g. the proton and the neutron in the framework
of an effective Yukawa theory. Since $\tilde{\eta}$ is small, the
interaction is dominated by one $\sigma+S$ - exchange and the amplitude
of the process has to be determined before passing to the non-relativistic
limit in order to obtain the corresponding interaction potential.
As for $\epsilon$ introduced above, one defines $\eta$ by $\eta g=\tilde{\eta}g'$.
In the following, except in Section \ref{sub:Constraints-from-},
$\epsilon$ and $\eta$ will be used instead of $\tilde{\epsilon}$
and $\tilde{\eta}.$

In a similar way as in \cite{Cline:2012is}, the dark fermions $F$
and $G$ will bind to form neutral dark hydrogenoid atoms of Bohr
radius $a'_{0}=\frac{1}{\mu\alpha'}$, where $\mu$ is the reduced
mass of the $F-G$ system and $\alpha'=\frac{e'}{4\pi}$. In principle,
the galactic dark matter halo could be populated by these neutral
dark atoms as well as by a fraction of dark ions $F$ and $G$, but
ref. \cite{McDermott:2010pa} shows that supernovae shock waves will
evacuate millicharged dark ions from the disk and that galactic magnetic
fields will prevent them from re-entering unless $\epsilon<9\times10^{-12}$
($m_{F,G}/$GeV), which is far below the values that we will be interested
in in the following to explain the signals of the direct-dark-matter-search
experiments. Therefore, the signals will only be induced by the interactions
of the dark atoms with matter in the detectors.

\subsection{Constraints from $\Upsilon$ and $J/\psi$ disintegrations\label{sub:Constraints-from-}}

A direct consequence of the mass mixing term in \eqref{eq:4} is that
a certain fraction of $\sigma$'s can convert into $S$ scalars and
then evade in the dark sector. This can be seen in the disintegrations
of quarkonium states such as the $J/\psi$ meson and the $1S$ and
$3S$ resonances of the $\Upsilon$ meson. The studied and unseen
processes are generically represented by \begin{equation}
\begin{array}{ccccc}
Q\bar{Q} & \rightarrow & \sigma\bar{\sigma} & \rightarrow & S\bar{S}\\
Q\bar{Q} & \rightarrow & \gamma\sigma & \rightarrow & \gamma S\end{array}\label{eq:30}\end{equation}
 where $Q\bar{Q}=\Upsilon(1S),\Upsilon(3S)$ or $J/\psi(1S)$. Because
of the partity $-1$ of these states, the disintegration in two particles
of parity $+1$ is forbidden, and one hence avoids the constraints
from the first process. From \cite{Balest:1994ch}, \cite{Aubert:2008as}
and \cite{Insler:2010jw}, the $90\%$ C.L. upper limits on the branching
ratios of the second process are respectively \begin{equation}
\begin{array}{lcc}
B(\Upsilon(1S)\rightarrow\gamma S) & < & 5.6\times10^{-5}\\
B(\Upsilon(3S)\rightarrow\gamma S) & < & 15.9\times10^{-6}\\
B(J/\psi(1S)\rightarrow\gamma S) & < & 4.3\times10^{-6}\end{array}\label{eq:31}\end{equation}

In the limit where the momenta of the constituent quarks are nul ($p=(M_{Q\bar{Q}}/2,\vec{0})$,
where $M_{Q\bar{Q}}$ is the mass of the $Q\bar{Q}$ meson), we get
\begin{equation}
\frac{B(Q\bar{Q}\rightarrow\gamma S)}{B\left(Q\bar{Q}\rightarrow e^{+}e^{-}\right)}=\frac{2\beta}{\alpha}\frac{M_{Q\bar{Q}}\left(M_{Q\bar{Q}}^{2}-m_{S}^{2}\right)}{\left(M_{Q\bar{Q}}^{2}+2m_{e}^{2}\right)\sqrt{M_{Q\bar{Q}}^{2}-4m_{e}^{2}}}\frac{\tilde{\eta}^{2}\left(m_{\sigma}^{2}+m_{S}^{2}\right)^{2}}{\left(m_{S}^{2}-m_{\sigma}^{2}\right)^{2}}\label{eq:32}\end{equation}
 where $B\left(Q\bar{Q}\rightarrow e^{+}e^{-}\right)$ is the branching
ratio of the disintegration of $Q\bar{Q}$ into a positron-electron
pair, $\alpha=\frac{e^{2}}{4\pi}=\frac{1}{137}$ is the fine structure
constant, $\beta=\frac{g^{2}}{4\pi}=16.5$, and $m_{e}$ is the mass
of the electron. $B\left(Q\bar{Q}\rightarrow e^{+}e^{-}\right)=\left(2.38\pm0.11\right)\%,\,\left(2.03\pm0.20\right)\%$
and $\left(5.94\pm0.06\right)\%$ \cite{Amsler:2008zzb}, respectively
for $Q\bar{Q}=\Upsilon(1S),\Upsilon(3S)$ and $J/\psi(1S)$.

Putting together \eqref{eq:31} and \eqref{eq:32}, one gets allowed
regions for parameters $\tilde{\eta}$ and $m_{S}$ from processes
\eqref{eq:31}. But for the rather small values of $m_{S}$ considered
here, expression \eqref{eq:32} turns out to be independent of the
mass of the scalar particle and the most stringent constraint comes
from the disintegration of $J/\psi\left(1S\right)$ : \begin{equation}
\tilde{\eta}<1.2\times10^{-4}\label{eq:33}\end{equation}

\subsection{Interactions of $F$ and $G$ fermions with nucleons and electrons}

The kinetic and mass mixings introduced in the lagrangian of the model
give rise, in the non-relativistic limit, to interaction potentials
between the particles $F$ and $G$ and standard protons, neutrons
and electrons.

The kinetic $\gamma-\Gamma$ mixing induces a Coulomb interaction
with protons or electrons with a potential given by \begin{equation}
V_{C}\left(r\right)=\pm\frac{\epsilon\alpha}{r}\label{eq:5}\end{equation}
 where the plus sign is for the proton$-F$ and electron$-G$ couplings,
and the minus sign for the electron$-F$ and proton$-G$ interactions.

The $\sigma-S$ mass mixing gives rise, in the non-relativistic limit,
to the one $\sigma+S$ - exchange potential between $F$ and a nucleon
\begin{equation}
V_{M}\left(r\right)=-\frac{\eta\left(m_{\sigma}^{2}+m_{S}^{2}\right)\beta}{r}\left(\frac{e^{-m_{\sigma}r}-e^{-m_{S}r}}{m_{S}^{2}-m_{\sigma}^{2}}\right)\label{eq:6}\end{equation}
Note that in the limit $m_{S}\rightarrow m_{\sigma}$, expression
\eqref{eq:6} becomes $V_{M}\left(r\right)=-\frac{\eta m_{\sigma}\beta}{2}e^{-m_{\sigma}r}$,
although this particular case won't be considered in the following.

\section{Thermalization of dark $FG$ atoms in terrestrial matter\label{sec:Thermalization-of-dark}}

Because of the motion of the Earth (and of the Sun) through the galactic
dark matter halo, an effective wind of dark atoms hits the surface
of our planet. These dark atoms penetrate the surface and undergo
elastic collisions with terrestrial atoms, and lose part of their
energy at each collision. If the number of collisions and the elastic-diffusion
cross section are sufficiently large, then the dark atoms can deposit
all their energy in the terrestrial matter before going out on the
other side of the Earth, or even thermalize between the surface and
an underground detector. The diffusions can be of two types : electromagnetic
(atom - dark atom) and $\sigma+S-$exchange (nucleus -$F$), from
potentials \eqref{eq:5} and \eqref{eq:6}. In the following, we shall
consider the terrestrial surface as made of {}``average'' atoms
of silicon, with atomic and mass numbers $Z_{m}=14$ and $A_{m}=28$
and mass $m_{m}=A_{m}m_{p}$, where $m_{p}$ is the mass of the proton.
The nuclear radius will be neglected here, since it is much smaller
than the wavelength of the incident particles at these energies, and
has therefore no influence on the elastic cross section.

\subsection{Interaction of dark $FG$ atoms with terrestrial atoms}

We assume that $m_{F}\gg m_{G}$, and hence that $m_{FG}\simeq m_{F}$,
where $m_{FG}$ is the mass of an $FG$ dark atom, so that in the
dark bound state $FG$, $F$ plays the role of a dark nucleus while
$G$ is spherically distributed around it. In this context, the dark
$FG$ atoms, as well as the terrestrial ones, are assimilated to uniformly
charged spheres of charges $-\epsilon e$ and $-Z_{m}e$ and radii
$a_{0}'$ and $a_{0}$, representing the respective electronic clouds,
with opposite point-like charges at their centers, corresponding to
the respective $F$ and silicon nuclei. Because the elastic interaction
cross section of a dark atom with a terrestrial atom has to be large
enough to allow thermalization before reaching an underground detector,
the atomic size of a dark atom will be of the same order as a standard
one. We take $1\,\textrm{\AA}$ as a reference for the atomic size
and set $a_{0}'=\frac{1}{m_{G}\alpha'}=a_{0}=1\,\lyxmathsym{\AA}$.
In view of the suggestion $e'\simeq e$ of Section \ref{sec:The-model}
, this gives $m_{G}\simeq m_{e}$.

We then obtain the atom - dark atom electrostatic interaction potential
as : \begin{equation}
\begin{array}{ccl}
V_{at} & = & \frac{\epsilon Z_{m}\alpha}{160a_{0}^{6}}\left(-r^{5}+30a_{0}^{2}r^{3}+80a_{0}^{3}r^{2}-288a_{0}^{5}+\frac{160a_{0}^{6}}{r}\right),\,\,\,\,\,\,\,\,\,\,\,\,\,\,\,\, r<a_{0}\\
 & = & \frac{\epsilon Z_{m}\alpha}{160a_{0}^{6}}\left(-r^{5}+30a_{0}^{2}r^{3}-80a_{0}^{3}r^{2}+192a_{0}^{5}-\frac{160a_{0}^{6}}{r}\right),\,\, a_{0}<r<2a_{0}\\
 & = & 0,\,\,\,\,\,\,\,\,\,\,\,\,\,\,\,\,\,\,\,\,\,\,\,\,\,\,\,\,\,\,\,\,\,\,\,\,\,\,\,\,\,\,\,\,\,\,\,\,\,\,\,\,\,\,\,\,\,\,\,\,\,\,\,\,\,\,\,\,\,\,\,\,\,\,\,\,\,\,\,\,\,\,\,\,\,\,\,\,\,\,\,\,\,\,\,\,\,\,\,\,\,\,\,\,\,\,\,\,\,\,\,\,\,\,\,\,\,\,\,\,\,\,\,\,\,\,\,\,\,\,\,\,\,\,\,\,\,\,\, r>2a_{0}\end{array}\label{eq:27}\end{equation}
 $r$ being the distance between both nuclei and $"at"$ standing
for $"atomic"$.

\begin{figure}
\begin{centering}
\includegraphics[scale=0.75]{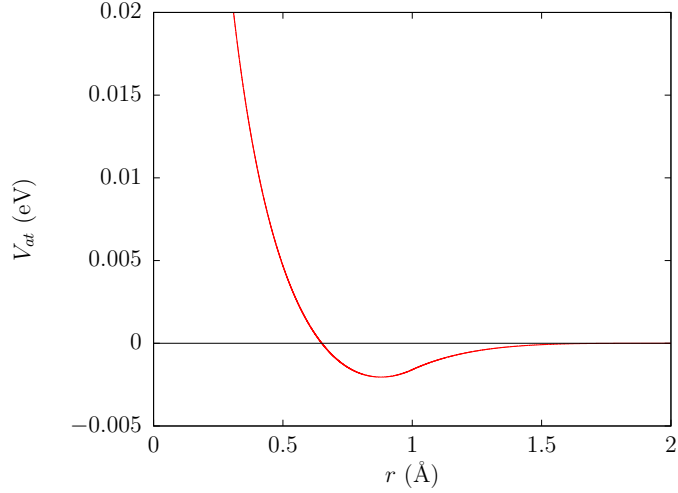}
\par\end{centering}

\caption{Shape of the silicon - $FG$ interaction potential $V_{at}$(eV) as
a function of the distance between nuclei $r$ ($\textrm{\AA}$),
with the best fit value $\epsilon=6.7\times10^{-5}$.\label{fig:pot_at}}

\end{figure}

The shape of $V_{at}$ is represented in Figure \ref{fig:pot_at}
for a silicon atom and for the best fit value of the kinetic mixing
parameter $\epsilon=6.7\times10^{-5}$, discussed in Section \ref{sec:Results}.
It shows a very shallow potential well at $r\simeq a_{0}$. Its depth,
of the order of $10^{-3}$ eV, doesn't allow to create atom - dark
atom bound states, as they would be destroyed by thermal excitation
in the Earth, where $T\sim300$ K corresponds to thermal energies
of the order of $10^{-2}$ eV. At smaller distance, when $r\lesssim0.6\textrm{\,\AA}$,
the Coulomb repulsion between nuclei starts to dominate. Thus no atomic
bound state can form with elements between the surface and an underground
detector.

In addition to this atom-dark atom interaction, both nuclei interact
through $\sigma+S-$exchange, corresponding to the potential \eqref{eq:6}
multiplied by the number of nucleons in a silicon nucleus: \begin{equation}
V_{nucl}\left(r\right)=-\frac{\eta\left(m_{\sigma}^{2}+m_{S}^{2}\right)A_{m}\beta}{r}\left(\frac{e^{-m_{\sigma}r}-e^{-m_{S}r}}{m_{S}^{2}-m_{\sigma}^{2}}\right)\label{eq:34}\end{equation}
 where $"nucl"$ stands for $"nuclear"$. Because $m_{\sigma}\gg m_{S}$,
this potential is very similar to a pure Yukawa potential $\sim-\frac{1}{r}e^{-mr}$.
Although it creates a deeper attractive well at short distance (of
the order of $m_{S}^{-1}\simeq100$ fm), this narrower potential will
neither admit stable bound states with the relatively light nuclei
present in terrestrial matter. Therefore, the interactions of $FG$
dark atoms in the Earth can be considered as purely elastic.

\subsection{Elastic diffusion cross section\label{sub:Elastic-diffusion-cross}}

The elastic differential cross sections corresponding to the potentials
\eqref{eq:27} and \eqref{eq:34} can be obtained by evaluating the
square of the modulus of the diffusion amplitude in the framework
of the Born approximation in the center of mass frame of the nucleus
- $F$ system : \begin{equation}
\begin{array}{cll}
\left(\frac{d\sigma}{d\Omega}\right)_{at} & = & \frac{\mu^{2}\epsilon^{2}Z_{m}^{2}\alpha^{2}}{a_{0}^{12}}\frac{1}{K^{16}}I^{2}\end{array}\label{eq:20}\end{equation}
 with \[
\begin{array}{cll}
I & = & 9\left(K^{2}a_{0}^{2}+1\right)+9\cos\left(2Ka_{0}\right)\left(K^{2}a_{0}^{2}-1\right)+12\cos\left(Ka_{0}\right)K^{4}a_{0}^{4}\\
\\ &  & -18\sin\left(2Ka_{0}\right)Ka_{0}-12\sin\left(Ka_{0}\right)K^{3}a_{0}^{3}+2K^{6}a_{0}^{6}\end{array}\]
 and \begin{equation}
\begin{array}{ccc}
\left(\frac{d\sigma}{d\Omega}\right)_{nucl} & = & 4\mu^{2}\eta^{2}A_{m}^{2}\beta^{2}\left(\frac{m_{\sigma}^{2}+m_{S}^{2}}{m_{S}^{2}-m_{\sigma}^{2}}\right)^{2}\left[\frac{1}{m_{\sigma}^{2}+K^{2}}-\frac{1}{m_{S}^{2}+K^{2}}\right]^{2}\end{array}\label{eq:35}\end{equation}
where $K=2k\sin\theta/2$ and $k=\sqrt{2\mu E}$ are the transferred
and initial momenta. $\theta$ is the deflection angle with respect
to the collisional axis and $\mu=\frac{m_{F}m_{m}}{m_{F}+m_{m}}$
is the reduced mass of the nucleus - $F$ system.

The total differential cross section corresponding to $V_{at}+V_{nucl}$
is finally given by the sum of \eqref{eq:20} and \eqref{eq:35} without
forgetting the interference term : \begin{equation}
\begin{array}{ccc}
\left(\frac{d\sigma}{d\Omega}\right)_{tot} & = & \left(\frac{d\sigma}{d\Omega}\right)_{at}+\left(\frac{d\sigma}{d\Omega}\right)_{nucl}-\frac{4\mu^{2}\epsilon\eta Z_{m}A_{m}\alpha\beta}{a_{0}^{6}}\left(\frac{m_{\sigma}^{2}+m_{S}^{2}}{m_{S}^{2}-m_{\sigma}^{2}}\right)\frac{I}{K^{8}}\left[\frac{1}{m_{\sigma}^{2}+K^{2}}-\frac{1}{m_{S}^{2}+K^{2}}\right]\end{array}\label{eq:36}\end{equation}

\subsection{Energy loss per unit path length : $\frac{dE}{dx}$}

At each collision with an atom of the terrestrial surface, a dark
atom loses an energy $\triangle K=\frac{p^{2}\left(\cos\theta-1\right)}{m_{m}}$
in the frame of the Earth, where $p$ is the relative momentum. The
energy loss per unit length in the frame of the Earth is then obtained
by integrating over all diffusion angles \begin{equation}
\frac{dE}{dx}=n_{m}\int_{\Omega}\triangle K\left(\frac{d\sigma}{d\Omega}\right)_{tot}d\Omega\label{eq:21}\end{equation}
 where $n_{m}$ is the numerical density of terrestrial atoms.

Of course, the linear path approximation is valid only when $m_{F}\gg m_{m}$,
but it gives in the other cases an upper limit on the penetration
length of the dark atoms through the Earth, which is of interest here.
To obtain it, one just needs to integrate the inverse of \eqref{eq:21}
from the initial energy of the dark atoms $E_{0}$ to the thermal
energy of the medium $E_{th}=\frac{3}{2}T_{m}$, where $T_{m}$ is
the temperature \begin{equation}
x=\int_{E_{th}}^{E_{0}}\frac{dE}{\left|dE/dx\right|}\label{eq:22}\end{equation}

\subsection{Penetration at a depth of $1$ km \label{sub:Penetration-at-a}}

\begin{figure}
\begin{centering}
\includegraphics[scale=0.75]{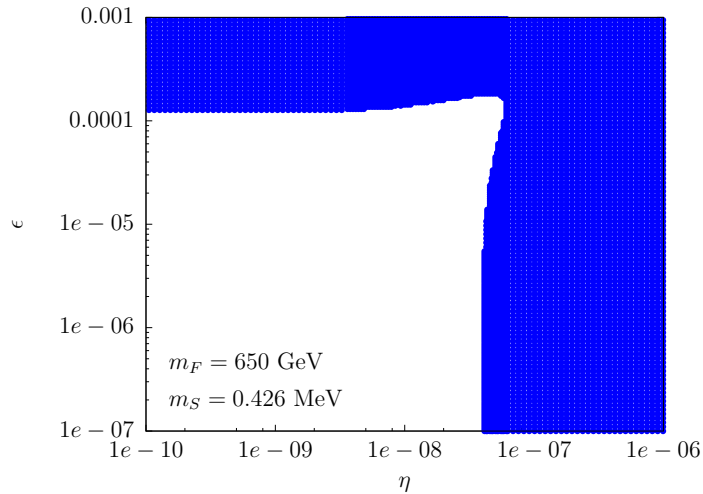}
\par\end{centering}

\caption{Region of parameters $\epsilon$ and $\eta$ (blue) where thermalization
of dark atoms occurs before reaching $1$ km underground, for the
best fit parameters $m_{F}=650$ GeV and $m_{S}=0.426$ MeV obtained
in Section \ref{sec:Results}.\label{fig:Region-of-parameters}}

\end{figure}

Figure \ref{fig:Region-of-parameters} shows the region (in blue)
of mixing parameters $\epsilon$ and $\eta$ where $x\leq1$ km ,
$1$ km being the typical depth at which underground detectors are
located, for the best fit values $m_{F}=650\,$GeV and $m_{S}=0.426$
MeV obtained in Section \ref{sec:Results}. In the blue region, thermalization
occurs before reaching $1$ km, while outside the dark atoms hit the
detector with non-thermal energies and can cause nuclear recoils.
The best-fit model, characterized by $m_{F}=650$ GeV, $m_{S}=0.426\,$MeV,
$\epsilon=6.7\times10^{-5}$ and $\eta=2.2\times10^{-7}$ clearly
satisfies the condition with $x\simeq40$ m.

Some interesting features are present in Figure \ref{fig:Region-of-parameters}
. At low $\eta$ ($\eta\lesssim10^{-9}$), thermalization is realized
entirely by the electromagnetic atom - dark atom interaction $V_{at}$,
for sufficiently large $\epsilon$ ($\epsilon\gtrsim10^{-4})$. When
$\eta$ increases ($10^{-9}\lesssim\eta\lesssim3\times10^{-8}$),
the limit on $\epsilon$ slightly increases. This conter-intuitive
behavior is due to the negative interference term present in the total
elastic cross section \eqref{eq:36} that increases with $\eta$.
For a certain range of $\eta$ ($3\times10^{-8}\lesssim\eta\lesssim6\times10^{-8}$),
$3$ regimes are visible : the first at low $\epsilon$, where thermalization
is mostly ensured by the nuclear interaction; the second at intermediate
$\epsilon$, where thermalization before $1$ km is not possibe because
the interference term partly compensates $\left(\frac{d\sigma}{d\Omega}\right)_{at}$
and $\left(\frac{d\sigma}{d\Omega}\right)_{nucl}$ in \eqref{eq:36};
the third at higher $\epsilon$, where thermalization is dominated
by $V_{at}$. Finally, at higher $\eta$ ($\eta\gtrsim6\times10^{-8}$),
all values of $\epsilon$ are possible, meaning that nuclear interaction
alone would be sufficient to thermalize.

\section{Interactions in underground detectors\label{sec:Interactions-in-underground}}

The dark atoms thermalize by elastic collisions in terrestrial matter
between the surface and the underground detector. Once they reach
thermal energies, they start drifting towards the center of the earth
until they reach the detector, where they undergo collisions with
the atoms of the active medium. Because of the Coulomb barrier due
to the repulsion between nuclei (seen in Figure \ref{fig:pot_at}
at $r\lesssim0.6\,\textrm{\AA}$), most of these collisions are elastic
but sometimes tunneling through the barrier can occur and bring a
dark nucleus $F$ into the region of the potential well present at
smaller distance, due to the exchange of $\sigma$ and $S$ between
$F$ and the nuclei of the detector. There, E1 transitions produce
de-excitation of the system to low-energy bound states by emission
of photons that can be detected, causing the observed signal. In the
following, only the part of the potential that is relevant for the
capture process is considered, i.e. the region $0<r\lesssim0.6\,\textrm{\AA}$,
where the interaction is dominated by the exchanges between $F$ and
the nucleus. The long-range part of the potential, $10^{3}$ to $10^{4}$
times smaller, does not affect the initial diffusion eigenstate and
the final bound state of the process and is therefore neglected, and
the dilute electronic and $G$ distributions, mostly transparent to
each other, follow passively their respective nuclei.

\subsection{Interactions of fermions $F$ with nuclei}

Because of their interactions with nucleons, the dark particles $F$
interact with nuclei. If a nucleus $N$ of mass number $A$ and atomic
number $Z$ is seen as a uniformly charged sphere of radius $R=r_{0}A^{1/3}$,
the integration of expressions \eqref{eq:5} and \eqref{eq:6} over
its electric and nuclear charge distributions gives \begin{equation}
\begin{array}{ccl}
V_{C}^{N}\left(r\right) & = & \frac{\epsilon Z\alpha}{2R}\left(3-\frac{r^{2}}{R^{2}}\right),\,\,\,\, r<R\\
 & = & \frac{\epsilon Z\alpha}{r},\,\,\,\,\,\,\,\,\,\,\,\,\,\,\,\,\,\,\,\,\,\,\,\,\,\,\,\,\,\,\, r>R\end{array}\label{eq:7}\end{equation}
 for the Coulomb potential, and \begin{equation}
\begin{array}{ccl}
V_{M}^{N}\left(r<R\right) & = & -\frac{V_{0}}{r}\left[2r\left(m_{\sigma}^{-2}-m_{S}^{-2}\right)+\left(R+m_{\sigma}^{-1}\right)m_{\pi}^{-2}\left(e^{-m_{\sigma}r}-e^{m_{\sigma}r}\right)e^{-m_{\sigma}R}\right.\\
\\ &  & \left.-\left(R+m_{S}^{-1}\right)m_{S}^{-2}\left(e^{-m_{S}r}-e^{m_{S}r}\right)e^{-m_{S}R}\right]\\
\\V_{M}^{N}\left(r>R\right) & = & -\frac{V_{0}}{r}\left[m_{\sigma}^{-2}e^{-m_{\sigma}r}\left(e^{m_{\sigma}R}\left(R-m_{\sigma}^{-1}\right)+e^{-m_{\sigma}R}\left(R+m_{\sigma}^{-1}\right)\right)\right.\\
\\ &  & \left.-m_{S}^{-2}e^{-m_{S}r}\left(e^{m_{S}R}\left(R-m_{S}^{-1}\right)+e^{-m_{S}R}\left(R+m_{S}^{-1}\right)\right)\right]\end{array}\label{eq:8}\end{equation}
 for the one $\sigma+S$ - exchange potential between $F$ and a nucleus.
In expression \eqref{eq:8}, $V_{0}=3\eta\left(m_{\sigma}^{2}+m_{S}^{2}\right)\beta/\left(2r_{0}^{3}\left(m_{S}^{2}-m_{\sigma}^{2}\right)\right)$,
where $r_{0}=1.2$ fm.

\begin{figure}
\begin{centering}
\includegraphics[scale=0.55]{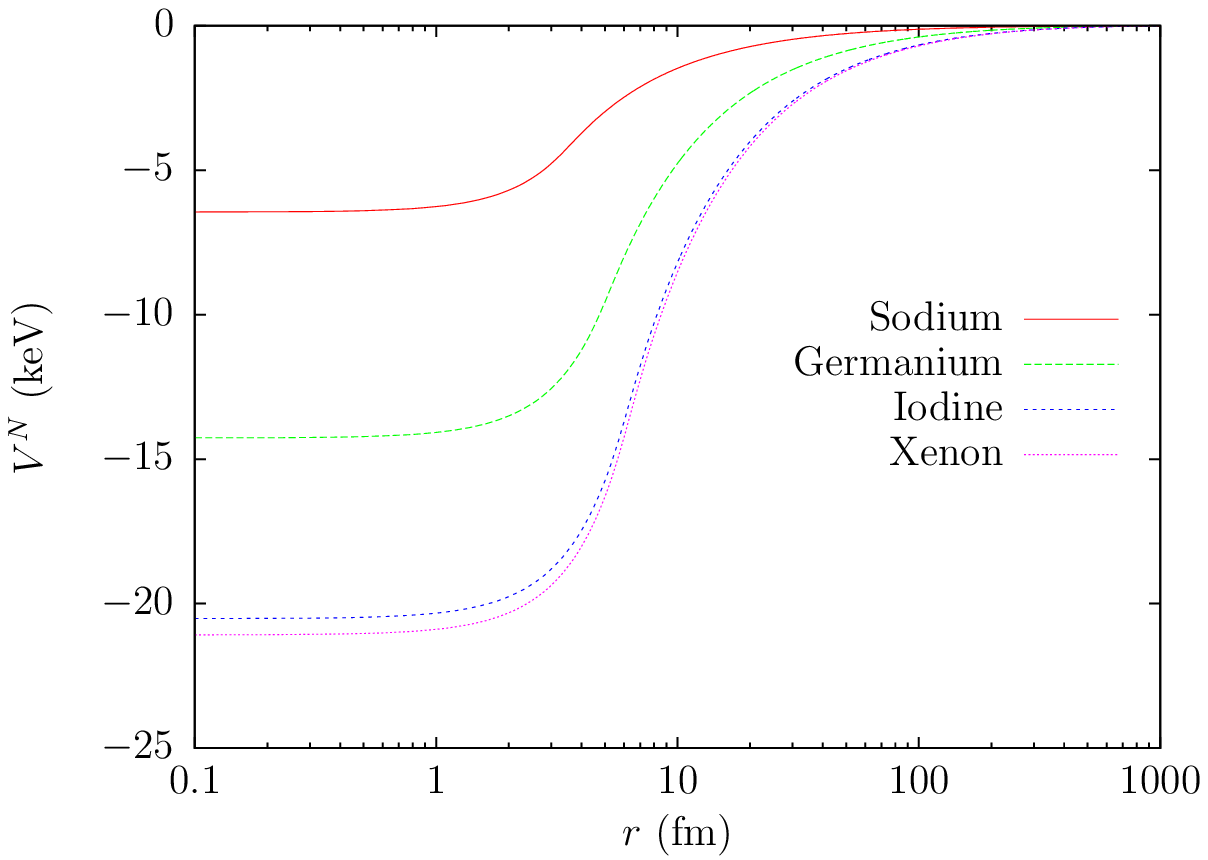}\includegraphics[scale=0.55]{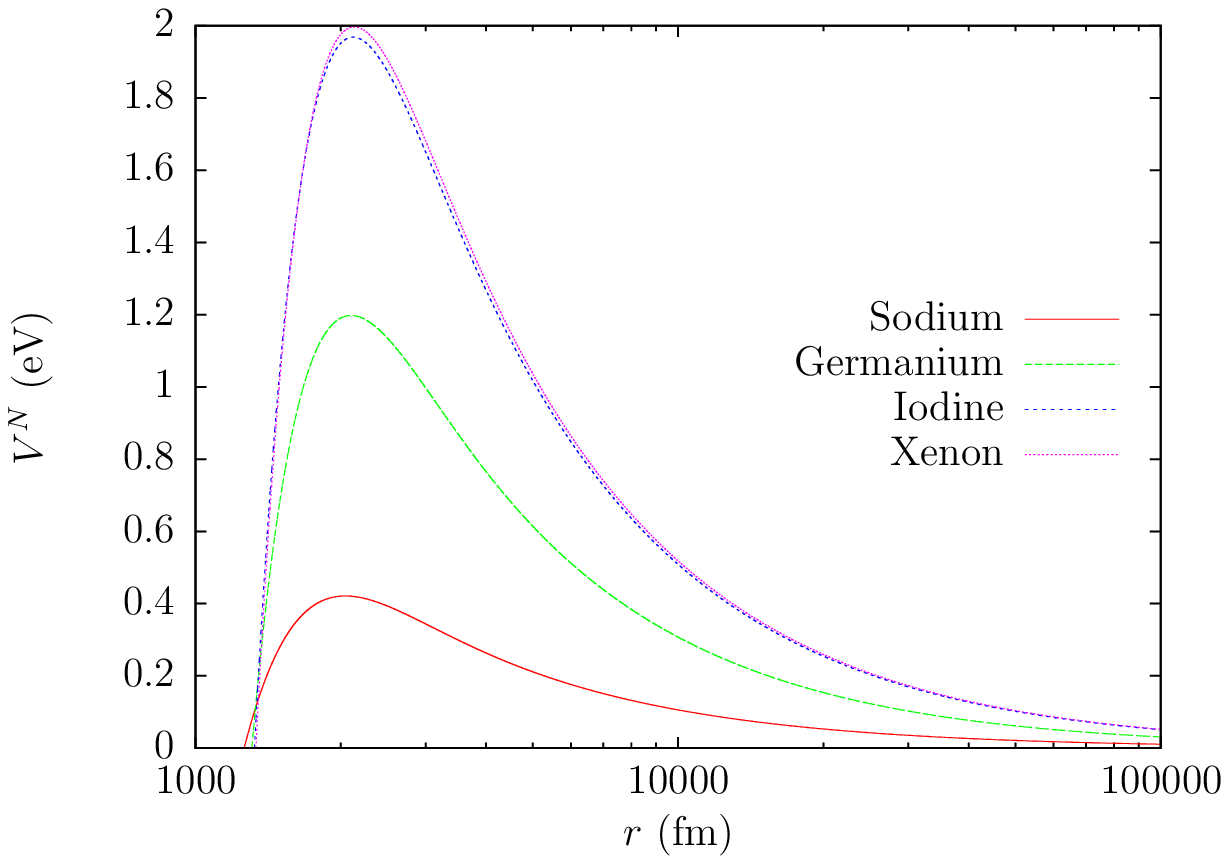}
\par\end{centering}

\caption{Shape of the total nucleus - $F$ interaction potential for light
(solid red), intermediate (long dashed green) and heavy (short dashed
blue, dotted magenta) nuclei consituting underground detectors. The
attractive part (nuclear well) is on the left (keV) and the repulsive
region (Coulomb barrier) is on the right (eV). The prefered parameters
of Section \ref{sec:Results} have been used.\label{fig:pot_nucleus}}

\end{figure}

Figure \ref{fig:pot_nucleus} shows the shape of the total potential
$V^{N}=V_{C}^{N}+V_{M}^{N}$ for light, intermediate and heavy nuclei,
all involved in underground detectors : Sodium (DAMA/LIBRA), Germanium
(CoGeNT, CDMS-II), Iodine (DAMA/LIBRA) and Xenon (XENON100). All these
potentials exhibit a Coulomb barrier, then an attractive well at shorter
distance. The height of the barrier as well as the depth and the width
of the well are determined by the values of the parameters $\epsilon$,
$\eta$ and $m_{S}$, taken here equal to the prefered values of Section
\ref{sec:Results}, but also depend on the nucleus. Typically, the
depth of the well is of several keV and the Coulomb barrier goes up
to several eV with a maximum being localized at about $2000$ fm.

\subsection{Bound-state formation mechanism}

At thermal energies, to order $v/c$, only the partial $s$-wave of
an incident plane wave on an attractive center is affected by the
potential. Considering the center-of-mass frame of the nucleus - $F$
system, this means that the largest contribution to tunneling corresponds
to tunneling through the Coulomb barrier at zero relative angular
momentum $l$. Due to selection rules, E1 transitions to final bound
states at $l=0$ are forbidden. It can also be shown that M1 and E2
transitions to such final levels are not present \cite{Segre:1977},
leaving only the possibility of captures of the particles $F$ in
two steps, i.e. first to levels at $l=1$ after tunneling and then
to levels at $l=0$, each one corresponding to an E1 transition. The
radiative capture of thermal particles $F$ therefore requires the
existence of bound states at least up to $l=1$ in the potential wells
of Figure \ref{fig:pot_nucleus}.

The transition probability per unit time for an electric multipole
radiation of order $q$ is given by \cite{Segre:1977} \begin{equation}
\lambda(q,m)=\frac{8\pi(q+1)}{q\left[(2q+1)!!\right]}\omega^{2q+1}\left|Q_{qm}\right|^{2}\label{eq:25}\end{equation}
 where $m=-q,...,q$, $\omega$ is the angular frequency of the emitted
radiation and the matrix element $Q_{qm}=e\sum_{j=1}^{N}\int r_{j}^{q}Y_{q}^{m*}\left(\theta_{j},\varphi_{j}\right)\psi_{f}^{*}\psi_{i}d\overrightarrow{r}$.
The sum is over all the electric charges $e_{j}$ of the system and
the spherical harmonics $Y_{q}^{m}$ are evaluated at the positions
of each of them. $\psi_{i}$ and $\psi_{f}$ are respectively the
initial and final states of the transition.

In the framework of this model, one has for the E1 capture from an
$s$ - state in the continuum to a bound $p$ - state, expressed in
the center-of-mass frame of the nucleus - $F$ system in terms of
relative coordinates $\overrightarrow{r}=\overrightarrow{r_{F}}-\overrightarrow{r_{N}}$
: \begin{equation}
\begin{array}{ccl}
\lambda\left(1,m\right) & = & \frac{16\pi}{9}\omega^{3}\left|Q_{1m}\right|^{2}\\
Q_{1m} & = & Ze\left(\frac{m_{F}}{m_{F}+m}\right)\int rY_{1}^{m*}\left(\theta,\varphi\right)\psi_{f}^{*}\left(\overrightarrow{r}\right)\psi_{i}\left(\overrightarrow{r}\right)d\overrightarrow{r}\end{array}\label{eq:28}\end{equation}
 where $m$ is the mass of the nucleus. The term in $Q_{1m}$ due
to the millicharged dark ion $F$ has been neglected with respect
to the term of the nucleus because of the factor $\epsilon$, that
brings a factor $\epsilon^{2}$ in the transition probability. The
initial and final states are expressed as \begin{equation}
\begin{array}{lll}
\psi_{i}\left(\vec{r}\right) & = & \frac{1}{k}R\left(r\right)\\
\psi_{f}\left(\overrightarrow{r}\right) & = & R_{f}\left(r\right)Y_{1}^{-1,0,1}\left(\theta,\varphi\right)\end{array}\label{eq:29}\end{equation}
 $R$ and $R_{f}$ being respectively the radial parts of the eigenfunctions
of the system at relative angular momenta $l=0$ and $l=1$, corresponding
to energies $E$ (positive, incident) and $E_{f}$ (negative, lowest
bound energy level at $l=1$) in the center-of-mass frame. $k=\sqrt{2\mu E}$,
where $\mu$ is the reduced mass of the nucleus - $F$ system, is
the momentum of the incident plane wave. The factor $\frac{1}{k}$
comes from the decomposition of a plane wave into partial waves.

The link between the transition probability $\lambda\left(1,m\right)$
and the capture cross section $\sigma_{capt}\left(1,m\right)$ is
made via the relation $\lambda\left(1,m\right)=n\sigma_{capt}\left(1,m\right)v$,
where $n$ is the number density of incident particles and $v=\left|\overrightarrow{v_{F}}-\overrightarrow{v_{N}}\right|$
is the relative velocity. $\psi_{i}$ is normalized in such a way
that there is one incident particle per unit volume ($n=1$), by numerically
solving the radial Schrodinger equation at $l=0$ for the positive
energy $E$ and matching the function $R\left(r\right)$ with the
asymptotically free amplitude. The total E1 capture cross section
$\sigma_{capt}$ is then obtained by summing the cross sections corresponding
to the three possible values of $m$ and one finally gets \begin{equation}
\sigma_{capt}=\frac{32\pi^{2}Z^{2}\alpha}{3\sqrt{2}}\left(\frac{m_{F}}{m_{F}+m}\right)^{2}\frac{1}{\sqrt{\mu}}\frac{\left(E-E_{f}\right)^{3}}{E^{3/2}}D^{2}\label{eq:9}\end{equation}
 where $D=\int_{0}^{\infty}rR_{f}\left(r\right)R\left(r\right)r^{2}dr$
and $\mu=\frac{m_{F}m}{m_{F}+m}$ is the reduced mass of the $F$
- nucleus system. $R_{f}$ and $E_{f}$ are obtained by solving the
radial Schrodinger equation at $l=1$ with the WKB approximation and
$R_{f}$ is normalized by demanding that $\int R_{f}^{2}(r)r^{2}dr=1$.

\subsection{Event counting rate}

In the active medium of a detector made of nuclei $N$ at temperature
$T$, both $F$ and $N$ have velocity distributions $P_{F}\left(\overrightarrow{v_{F}}^{lab}\right)$
and $P_{N}\left(\overrightarrow{v_{N}}^{lab}\right)$, where {}``$lab$''
stands for {}``laboraty frame'' . We take them of the same Maxwellian
form \begin{equation}
\begin{array}{ccccc}
P_{F}\left(\overrightarrow{v_{F}}^{lab}\right) & = & P\left(\overrightarrow{v_{F}}^{lab}\right) & = & \left(\frac{m_{F}}{2\pi T}\right)^{3/2}e^{-m_{F}v_{F}^{lab^{2}}/2T}\\
P_{N}\left(\overrightarrow{v_{N}}^{lab}\right) & = & P\left(\overrightarrow{v_{N}}^{lab}\right) & = & \left(\frac{m}{2\pi T}\right)^{3/2}e^{-mv_{N}^{lab^{2}}/2T}\end{array}\label{eq:10}\end{equation}

The event counting rate $R$ per unit volume of the detector is given
by \begin{equation}
R=n_{F}n_{N}<\sigma_{capt}v>\label{eq:11}\end{equation}
 where $n_{F}$ and $n_{N}$ are the numerical densities of $F$ and
$N$ in the detector -and $<\sigma_{capt}v>$ is the thermally averaged
capture cross section times the relative velocity \begin{equation}
<\sigma_{capt}v>=\int\sigma_{capt}vP\left(\overrightarrow{v_{F}}^{lab}\right)P\left(\overrightarrow{v_{N}}^{lab}\right)d^{3}v_{F}^{lab}d^{3}v_{N}^{lab}\label{eq:12}\end{equation}

Passing to center-of-mass and relative velocities $\overrightarrow{v}_{CM}$
and $\overrightarrow{v}$, using \eqref{eq:9}, \eqref{eq:10}, \eqref{eq:11},
\eqref{eq:12} and performing the integration over the center-of-mass
variables, we get \begin{equation}
R=8n_{F}n_{N}\frac{1}{\left(2\pi T\right)^{3/2}}\frac{1}{\mu^{1/2}}\int_{0}^{\infty}\sigma_{capt}\left(E\right)Ee^{-E/T}dE\label{eq:13}\end{equation}
 where $E=\frac{1}{2}\mu v^{2}$ is the total energy in the center-of-mass
frame. 

Considering the annual modulation scenario and requiring that the
density of particles $F$ in the detector is determined by the equilibrium
between the incoming flux at the terrestrial surface and the down-drifting
thermalized flux, driven by gravity, one can write down the numerical
density $n_{F}$ within the detector as a function modulated in time
: \begin{equation}
\begin{array}{ccc}
n_{F} & = & n_{F}^{0}+n_{F}^{m}\cos\left(\omega\left(t-t_{0}\right)\right)\end{array}\label{eq:14}\end{equation}
 where $\omega=\frac{2\pi}{T_{orb}}$ is the angular frequency of
the orbital motion of the Earth around the Sun and $t_{0}\simeq$
June 2 is the period of the year when the Earth and Sun orbital velocities
are aligned. The constant part is given by \begin{equation}
n_{F}^{0}=\frac{n_{0}\, n\,\left\langle \sigma_{at}v\right\rangle }{4g}V_{h}\label{eq:15}\end{equation}
 while the annual modulation of the concentration is characterized
by the amplitude \begin{equation}
n_{F}^{m}=\frac{n_{0}\, n\,\left\langle \sigma_{at}v\right\rangle }{4g}V_{E}\cos\gamma\label{eq:16}\end{equation}
 $V_{h}=220\times10^{5}$ cm/s is the orbital velocity of the Sun
around the galactic center, $V_{E}=29.5\times10^{5}$ cm/s is the
Earth orbital velocity around the sun, $\gamma\simeq60^{\circ}$ is
the inclination angle of the Earth orbital plane with respect to the
galactic plane, $n_{0}=\frac{3\times10^{-4}}{S_{3}}$ cm$^{-3}$ is
the local density of the dark atoms, $n\simeq5\times10^{22}$ cm$^{-3}$
is the numerical density of atoms in the terrestrial crust, $g=980$
cm/s$^{2}$ is the acceleration of gravity and $n\,\left\langle \sigma_{at}v\right\rangle $
is the rate of elastic collisions between a thermalized dark atom
$FG$ and terrestrial atoms. $\sigma_{at}$ is obtained by integrating
the differential cross section \eqref{eq:20} from section \ref{sub:Elastic-diffusion-cross}
over all diffusion angles in the case of a silicon atom and $v$ is
the relative velocity between a dark atom and a terrestrial atom.
Note that $\sigma_{at}$ dominates over $\sigma_{nucl}$ at low energies,
so there is no need to consider $\sigma_{tot}$ here.

Expression \eqref{eq:14} may be inserted into \eqref{eq:13} to get
an annually modulated counting rate per unit volume of the detector
\begin{equation}
R=R^{0}+R^{m}\cos\left(\omega\left(t-t_{0}\right)\right)\label{eq:17}\end{equation}

In counts per day and per kilogram (cpd/kg) of detector, the constant
and modulated parts of the signal will respectively be given by \begin{equation}
\begin{array}{ccc}
R^{0} & = & Cn_{F}^{0}\int_{0}^{\infty}\sigma_{capt}\left(E\right)Ee^{-E/T}dE\\
R^{m} & = & Cn_{F}^{m}\int_{0}^{\infty}\sigma_{capt}\left(E\right)Ee^{-E/T}dE\end{array}\label{eq:18}\end{equation}
 with \[
C=24.10^{10}\frac{QtN_{Av}}{M_{mol}}\frac{1}{(2\pi T)^{3/2}}\frac{1}{\mu^{1/2}}\]
where $Q=1000$ g, $t=86400$ s, $N_{Av}=6.022\times10^{23}$ and
$M_{mol}$ is the molar mass of the active medium of the detector
in g/mol.

\section{Results\label{sec:Results}}

The presented model intends to reproduce the positive results of direct
dark matter searches experiments, such as DAMA/LIBRA and CoGeNT, without
contradicting the negative results of some others, such as XENON100
or CDMS-II/Ge.

The DAMA/LIBRA experiment observes an integrated modulation amplitude
$\tilde{R}_{DAMA}^{m}=\left(0.0464\pm0.0052\right)$ cpd/kg in the
energy interval $\left(2-6\right)$ keV \cite{Bernabei:2010mq}, while
the temporal analysis of CoGeNT has given $\tilde{R}_{CoGeNT}^{m}=\left(1.66\pm0.38\right)$
cpd/kg in the interval $\left(0.5-2.5\right)$ keV \cite{Aalseth:2011wp}.

Here, in a first approximation and for simplicity, the signal is supposed
to be made of one monochromatic line of energy $\Delta E_{DAMA},\,\Delta E_{CoGeNT}$.
It would be very interesting to reproduce the observed energy distributions
of the rates by taking into account the possible transitions to the
different $s$ - states, but this is postponed to another paper.

One first solves the Schrodinger equation independent on time with
potential $V^{N}=V_{C}^{N}+V_{M}^{N}$ in cases of Iodine ($^{127}$I
component of DAMA/LIBA detector), Germanium ($^{74}$Ge component
of CoGeNT detector) and Xenon ($^{132}$Xe component of XENON100 detector)
with the WKB approximation. This gives good estimates of the eigenvalues
and eigenfunctions of the respective two-body bound state problems.
The bound eigenfunctions are normalized numercially before computing
the constant or modulated number density of $F$ particles \eqref{eq:15}
or \eqref{eq:16}. The constant or modulated part of the event rate
is finally computed for each nucleus from \eqref{eq:18} with the
expression \eqref{eq:9} of the capture cross section, at the operating
temperatures of the different detectors, i.e. $T=300,\,73$ and $173$
K for DAMA/LIBRA, CoGeNT and XENON100 respectively.

One set of parameters that reproduces the data well and the corresponding
transitions energies ($\Delta E$), lowest levels at $l=1$ ($E^{l=1}$)
and rates ($R^{0}$ and $R^{m}$) are given in Table \ref{tab:Best-fit-parameters}.

\begin{table}
\begin{centering}
\begin{tabular}{|c|c|c|c|c|}
\hline 
 & $m_{F}$ (GeV) & $m_{S}$ (MeV) & $\epsilon$ & $\eta$\tabularnewline
\hline 
Best fit & $650$ & $0.426$ & $6.7\times10^{-5}$ & $2.2\times10^{-7}$\tabularnewline
\hline
\hline 
 & $\Delta E$ (keV) & $E^{l=1}$ (keV) & \multicolumn{1}{c|}{$R^{0}$ (cpd/kg)} & $R^{m}$ (cpd/kg)\tabularnewline
\hline 
DAMA/LIBRA & $3.8$ & $-2.0$ & - & $0.045$\tabularnewline
\hline 
CoGeNT & $1.4$ & $-0.4$ & - & $1.673$\tabularnewline
\hline 
XENON100 & $4.1$ & $-2.3$ & $8.455\times10^{-5}$ & -\tabularnewline
\hline
\end{tabular}
\par\end{centering}

\caption{Best fit parameters and predicted transitions energies and event counting
rates for DAMA/LIBRA, CoGeNT and XENON100 experiments.\label{tab:Best-fit-parameters}}

\end{table}

The energies of the signals and the event rates are well reproduced
for the DAMA and CoGeNT experiments. The lowest levels at $l=1$ give
rise to E1 captures that emit photons at threshold ($2$ keV for DAMA)
or below threshold ($0.5$ keV for CoGeNT) and only the photon emitted
during the second E1 transition from a $p$ - state to an $s$ - state
is observed, making the captures look like single-hit events. The
low predicted rate for XENON100 corresponds, over the total exposure
of the experiment \cite{Aprile:2012nq}, to $\simeq0.6$ events. Therefore,
no dark matter event should have occured within the XENON100 detector,
which is consistent with observations. Also, if we set $g'=g$, so
that $\eta=\tilde{\eta},$ the best fit value of $\eta$ is well below
the limit \eqref{eq:33} obtained from vector meson disintegrations.

Computing the penetration length \eqref{eq:22} with the parameters
of Table \ref{tab:Best-fit-parameters}, one finds that the dark atoms
thermalize after $\simeq40$ m, so that they reach the detectors at
thermal energies, as required by the model and already announced in
Subsection \ref{sub:Penetration-at-a}.

In a cooled detector, the dark atoms also have to thermalize when
they pass from the laboratory room to the active medium, i.e. at the
edge of the detector or over a distance smaller than its size. One
can roughly estimate the penetration in a detector with the same formula
\eqref{eq:22}, by setting $E_{0}=\frac{3}{2}T_{room}$ and $E_{th}=\frac{3}{2}T$,
even if here the motions of the atoms in the thermalizer should be
taken into account and the straight-line-path approximation is more
questionable. This gives, for CoGeNT and XENON100, penetration lengths
$\simeq1\,\textrm{\AA}$, which is clearly much smaller than the size
of the detectors and corresponds to thermalizations directly at the
edges.

This model predicts an event rate consistent with zero in any cryogenic
detector ($T\simeq1$ mK), due to the Coulomb barrier of the nucleus
- $F$ potential that prevents particles with very small energies
to be captured in the well. This is in agreement with the negative
results of the cryogenic CDMS-II/Ge (Germanium) experiment, in which
thermalization when entering the detector is realized after $\simeq1$
$\mu$m.

In the same manner, we predict no events in the cryogenic CDMS-II/Si
(Silicon) and CRESST-II detectors, in contradiction with the three
events recently observed by the former and the signal of the latter.
However, the penetration length in a cryogenic detector made of Silicon
as CDMS-II/Si is $\simeq1$ mm, i.e. $3$ orders of magnitude larger
than its equivalent in Germanium. This is essentially due to the smaller
electric charge of a Silicon nucleus, giving a weaker stopping power.
In this case, more collisions happen near the edge of the detector,
while the dark atoms are still at room temperature and hence more
likely to cross the Coulomb barrier. These peripheral collisions should
therefore be studied in detail to explain the events of some cryogenic
detectors.

In this analysis, attention has been paid to the Iodine component
of the DAMA detector, while it is constituted by a crystal of NaI,
and hence also of Sodium. Some part of the signal could come from
this other component, but it turns out that the only bound state with
$^{23}$Na is very shallow ($-61$ eV) and is at $l=0$. There is
therefore no $p$ - state on which the capture can happen, and the
signal of DAMA is due only to its Iodine component. One can try to
reproduce data directly with the Sodium component, but in that case
the levels obtained afterward with Iodine are much too low (because
the potential well is lower, as seen in Figure \ref{fig:pot_nucleus})
and give rise to a signal out of the detection interval of DAMA.

The fact that DAMA data are reproduced with the heavy component, Iodine,
and not with the light one, Sodium, is in fact an advantage of the
model, since in this situation, light isotopes do not have any bound
states with dark atoms. The first element presenting an $s$ bound
state is Oxygen ($Z=8$) while the first one having at least one $p$
bound state is Phosphorus ($Z=15$). Binding is therefore impossible
for very light nuclei with $Z\leq7$, preventing the formation of
anomalous isotopes during BBN, while heavy isotopes cannot form on
Earth with nuclei $Z\leq14$, representing the majority of terrestrial
elements.

\section{Conclusion\label{sec:Conclusion}}

We have presented a model in which a fraction of the dark matter density
($5\%$ or less) is realized by two new species of fermions $F$ and
$G$, forming hydrogenoid atoms with standard atomic size through
a dark $U(1)$ gauge interaction carried out by a dark massless photon.
Dark scalar particles $S$ are exchanged by the nuclei $F$ because
of a Yukawa coupling between $F$ and $S$. A kinetic photon - dark
photon mixing and a mass $\sigma$ - $S$ mixing, respectively characterized
by small dimensionless mixing parameters $\epsilon$ and $\eta$,
induce interactions between the dark sector and the ordinary one.
The dark atoms interact elastically in terrestrial matter until they
thermalize, in such a way that they reach underground detectors with
thermal energies. There, they form bound states with nuclei by radiative
capture, causing the emission of photons that create the observed
signals. The model reproduces well the positive results from DAMA/LIBRA
and CoGeNT, without contradicting the negative results from XENON100
with the following parameters : $m_{F}=650$ GeV, $m_{S}=0.426$ MeV,
$\epsilon=6.7\times10^{-5}$ and $\eta=2.2\times10^{-7}$. It naturally
prevents any signal in a cryogenic detector ($T\sim1$ mK), which
is consistent with CDMS-II/Ge. Further studies have to be performed
to explain the presence of a signal in CRESST-II, and possibly in
CDMS-II/Si, especially by considering the collisions of the dark atoms
at the edge of the detector, when they are still at room temperature
while the detector is colder.

\section*{Acknowledgments}

I am grateful to my advisor, J.R. Cudell, for key reading suggestions
and many discussions concerning this work. My thanks go to M. Khlopov
for inspiring ideas and discussions and to M. Tytgat for useful comments.
I thank the Belgian Fund F.R.S.-FNRS, by which I am supported as a
Research Fellow.

\bibliographystyle{jd}
\nocite{*}
\bibliography{references}

\providecommand{\href}[2]{#2}\begingroup\raggedright\begin{thebibliography}{10}

\bibitem{Bernabei:2010mq}
{\bfseries DAMA/LIBRA} Collaboration, R.~Bernabei {\em et~al.}, ``{New results
  from DAMA/LIBRA},''
  \href{http://dx.doi.org/10.1140/epjc/s10052-010-1303-9}{{\em Eur. Phys. J.}
  {\bfseries C67} (2010) 39--49},
\href{http://arxiv.org/abs/1002.1028}{{\ttfamily arXiv:1002.1028
  [astro-ph.GA]}}.

\bibitem{Aalseth:2012if}
{\bfseries CoGeNT} Collaboration, C.~Aalseth {\em et~al.}, ``{CoGeNT: A Search
  for Low-Mass Dark Matter using p-type Point Contact Germanium Detectors},''
  \href{http://dx.doi.org/10.1103/PhysRevD.88.012002}{{\em Phys. Rev. D 88}
  {\bfseries 012002} (2013) },
\href{http://arxiv.org/abs/1208.5737}{{\ttfamily arXiv:1208.5737
  [astro-ph.CO]}}.

\bibitem{Angloher:2011uu}
G.~Angloher, M.~Bauer, I.~Bavykina, A.~Bento, C.~Bucci, {\em et~al.},
  ``{Results from 730 kg days of the CRESST-II Dark Matter Search},''
  \href{http://dx.doi.org/10.1140/epjc/s10052-012-1971-8}{{\em Eur. Phys. J.}
  {\bfseries C72} (2012) 1971},
\href{http://arxiv.org/abs/1109.0702}{{\ttfamily arXiv:1109.0702
  [astro-ph.CO]}}.

\bibitem{Agnese:2013rvf}
{\bfseries CDMS-II} Collaboration, R.~Agnese {\em et~al.}, ``{Dark Matter
  Search Results Using the Silicon Detectors of CDMS II},'' {\em Phys. Rev.
  Lett.} (2013) ,
\href{http://arxiv.org/abs/1304.4279}{{\ttfamily arXiv:1304.4279 [hep-ex]}}.

\bibitem{Aprile:2012nq}
{\bfseries XENON100} Collaboration, E.~Aprile {\em et~al.}, ``{Dark Matter
  Results from 225 Live Days of XENON100 Data},''
  \href{http://dx.doi.org/10.1103/PhysRevLett.109.181301}{{\em Phys. Rev.
  Lett.} {\bfseries 109} (2012) 181301},
\href{http://arxiv.org/abs/1207.5988}{{\ttfamily arXiv:1207.5988
  [astro-ph.CO]}}.

\bibitem{Ahmed:2010wy}
{\bfseries CDMS-II} Collaboration, Z.~Ahmed {\em et~al.}, ``{Results from a
  Low-Energy Analysis of the CDMS II Germanium Data},''
  \href{http://dx.doi.org/10.1103/PhysRevLett.106.131302}{{\em Phys. Rev.
  Lett.} {\bfseries 106} (2011) 131302},
\href{http://arxiv.org/abs/1011.2482}{{\ttfamily arXiv:1011.2482
  [astro-ph.CO]}}.

\bibitem{Foot:2012rk}
R.~Foot, ``{Mirror dark matter interpretations of the DAMA, CoGeNT and
  CRESST-II data},'' \href{http://dx.doi.org/10.1103/PhysRevD.86.023524}{{\em
  Phys. Rev.} {\bfseries D86} (2012) 023524},
\href{http://arxiv.org/abs/1203.2387}{{\ttfamily arXiv:1203.2387 [hep-ph]}}.

\bibitem{Cline:2012is}
J.~M. Cline, Z.~Liu, and W.~Xue, ``{Millicharged Atomic Dark Matter},''
  \href{http://dx.doi.org/10.1103/PhysRevD.85.101302}{{\em Phys. Rev.}
  {\bfseries D85} (2012) 101302},
\href{http://arxiv.org/abs/1201.4858}{{\ttfamily arXiv:1201.4858 [hep-ph]}}.

\bibitem{Khlopov:2010ik}
M.~Y. Khlopov, A.~G. Mayorov, and E.~Y. Soldatov, ``{The dark atoms of dark
  matter},'' {\em Prespace. J.} {\bfseries 1} (2010) 1403--1417,
\href{http://arxiv.org/abs/1012.0934}{{\ttfamily arXiv:1012.0934
  [astro-ph.CO]}}.

\bibitem{Khlopov:2011me}
M.~Y. Khlopov, A.~G. Mayorov, and E.~Y. Soldatov, ``{Towards Nuclear Physics of
  OHe Dark Matter},'' in {\em {Proceedings to the $14^{th}$ Workshop "What
  Comes Beyond the Standard Model"}}, pp.~94--102.
\newblock 2011.
\newblock
\href{http://arxiv.org/abs/1111.3577}{{\ttfamily arXiv:1111.3577 [hep-ph]}}.
\newblock

\bibitem{Cudell:2012fw}
J.~Cudell, M.~Khlopov, and Q.~Wallemacq, ``{The nuclear physics of OHe},'' in
  {\em {Proceedings to the $15^{th}$ Workshop "What Comes Beyond the Standard
  Model"}}, pp.~10--27.
\newblock 2012.
\newblock
\href{http://arxiv.org/abs/1211.5684}{{\ttfamily arXiv:1211.5684
  [astro-ph.CO]}}.
\newblock

\bibitem{2002ApJ...564...60M}
J.~Miralda-Escud{\'e}, ``{A Test of the Collisional Dark Matter Hypothesis from
  Cluster Lensing},'' \href{http://dx.doi.org/10.1086/324138}{{\em
  Astrophysical Journal} {\bfseries 564} (2002) 60--64},
\href{http://arxiv.org/abs/0002050}{{\ttfamily arXiv:0002050 [astro-ph]}}.

\bibitem{Fan:2013yva}
J.~Fan, A.~Katz, L.~Randall, and M.~Reece, ``{Double-Disk Dark Matter},''
\href{http://arxiv.org/abs/1303.1521}{{\ttfamily arXiv:1303.1521
  [astro-ph.CO]}}.

\bibitem{Amsler:2008zzb}
{\bfseries Particle Data Group} Collaboration, C.~Amsler {\em et~al.},
  ``{Review of Particle Physics},''
\href{http://dx.doi.org/10.1016/j.physletb.2008.07.018}{{\em Phys. Lett.}
  {\bfseries B667} (2008) 1--1340}.

\bibitem{Erkol:2005jz}
G.~Erkol, R.~Timmermans, and T.~Rijken, ``{The Nucleon-sigma coupling constant
  in QCD Sum Rules},'' \href{http://dx.doi.org/10.1103/PhysRevC.72.035209}{{\em
  Phys. Rev.} {\bfseries C72} (2005) 035209},
\href{http://arxiv.org/abs/0603056}{{\ttfamily arXiv:0603056 [nucl-th]}}.

\bibitem{McDermott:2010pa}
S.~D. McDermott, H.-B. Yu, and K.~M. Zurek, ``{Turning off the Lights: How Dark
  is Dark Matter?},'' \href{http://dx.doi.org/10.1103/PhysRevD.83.063509}{{\em
  Phys. Rev.} {\bfseries D83} (2011) 063509},
\href{http://arxiv.org/abs/1011.2907}{{\ttfamily arXiv:1011.2907 [hep-ph]}}.

\bibitem{Balest:1994ch}
{\bfseries CLEO} Collaboration, R.~Balest {\em et~al.}, ``{$\Upsilon(1S) \to
  \gamma$ + noninteracting particles},''
\href{http://dx.doi.org/10.1103/PhysRevD.51.2053}{{\em Phys. Rev.} {\bfseries
  D51} (1995) 2053--2060}.

\bibitem{Aubert:2008as}
{\bfseries BaBar} Collaboration, B.~Aubert {\em et~al.}, ``{Search for
  Invisible Decays of a Light Scalar in Radiative Transitions $\Upsilon(3S) \to
  \gamma$ A0},''
\href{http://arxiv.org/abs/0808.0017}{{\ttfamily arXiv:0808.0017 [hep-ex]}}.

\bibitem{Insler:2010jw}
{\bfseries CLEO} Collaboration, J.~Insler {\em et~al.}, ``{Search for the Decay
  $J/\psi \to \gamma$ + invisible},''
  \href{http://dx.doi.org/10.1103/PhysRevD.81.091101}{{\em Phys. Rev.}
  {\bfseries D81} (2010) 091101},
\href{http://arxiv.org/abs/1003.0417}{{\ttfamily arXiv:1003.0417 [hep-ex]}}.

\bibitem{Segre:1977}
E.~Segre, {\em {Nuclei and Particles}}.
\newblock W. A. Benjamin, Inc., 2nd~ed., 1977.

\bibitem{Aalseth:2011wp}
C.~Aalseth, P.~Barbeau, J.~Colaresi, J.~Collar, J.~Diaz~Leon, {\em et~al.},
  ``{Search for an Annual Modulation in a P-type Point Contact Germanium Dark
  Matter Detector},''
  \href{http://dx.doi.org/10.1103/PhysRevLett.107.141301}{{\em Phys. Rev.
  Lett.} {\bfseries 107} (2011) 141301},
\href{http://arxiv.org/abs/1106.0650}{{\ttfamily arXiv:1106.0650
  [astro-ph.CO]}}.

\end{thebibliography}\endgroup

\end{document}